# 12 YEARS OF DATA, RESULTS AND EXPERIENCES IN THE EUROPEAN RESEARCHERS' NIGHT PROJECT

G. Mazzitelli[1,3], S. Arnone[1,3], M. Bramato[3], I. Capra[3], G. Ciocca[2,3], A. Della Ceca[2,3], R. Giovanditti[3], C. Grasso[3], D. Maselli[1,3], G. Sanzone[3], D. Sereni[3], F. Spagnoli[3]

[1]Istituto Nazionale di Fisica Nucleare Laboratori Nazionali di Frascati
Via E. Fermi 00044 Frascati, Rome, ITALY
[2]G.Eco - Gruppo didattica dell'Ecologia
Piazza delle Iris, 28, 00171, Rome, ITALY
[3]Frascati Scienza, Scuderie Aldobrandini,
Piazza Marconi 3 – 00044 Frascati, Rome, ITALY

## Abstract

In the last twelve years, the researchers in the Roman area have organized the European Researchers' Night (ERN), a project funded by the European Commission aiming to bring researchers closer to the general public and stimulating the interest in science careers among young people. Since the first edition in 2006, when the National Laboratory of INFN's in Frascati hosted 4000 visitors, the project has always grown up by reaching 50,000 attendees and involving more than 50 scientific partners in 2017. In the last edition, the Frascati Scienza association, which was set up to coordinate the event, has operated in over 30 Italian cities from south to north of the peninsula. In addition, the MADE IN SCIENCE project - European Research Week 2016/2017 - has been one of the largest projects funded by the European Commission in the framework of the European Researchers' Night and it is often referred as a positive model for events' organization and communication to the general public.

The twelve years of data collection and results obtained, as well as some of the most important experiences in public communication of science will be shown in this paper.

## 1 INTRODUCTION

The National Laboratory of Frascati (LNF), part of the Italian National Institute for Nuclear Physics (INFN), is one of the largest Italian research infrastructures located in the south area of Rome, close to the Frascati city, epicentre since the '50s of many research facilities belonging to the most important Italian research institutions, and operating at national and international level. Indeed, The Frascati research area can be considered one of the largest in Europe: the LNF hosts DAFNE, one of the five particle accelerators in the world aimed to discovery the constituents of our Universe; the Italian National Agency for New Technologies, Energy and Sustainable Economic Development (ENEA), which hosts the FTU (Frascati Tokamak Upgrade), one of the European facility to test fusion principle; the European Space Agency (ESA-ESRIN), coordinating all the European space activities; the Italian Space Agency (ASI), funding the Italian space research and data exploitation; the National Institute for Astrophysics (INAF), with an observatory and a research centre for plasma physics; and last but not least, the University of Tor Vergata and the ARTOV of the National Research Centre (CNR), they complete the framework in which 3000 researchers operate everyday. A suitable site where design a large outreach program.

In 2006, a small group of researchers, technicians and administrative staff of the LNF won the first European Researcher's Night (ERN) proposal [1], a one night-event, that falls every last Friday of September, to promote the researcher's figure and its work. The European Commission has started this program in the Marie Sklodowska Curie Action (MSCA) in 2005, under the PEOPLE specific programme by organising a first public event in Brussels, aimed to enhance researchers' public recognition, its role within the society, to stimulate youngsters and people at large to be involved in research and understand its impact in everyday life. Since the beginning, in 2006, ENEA, ESA and INAF joined the collaboration with INFN, as well as the Municipality of Frascati and the Cultural and Research Department of the Lazio Region, that has started to co-fund the initiative.

Today, after twelve editions, the project has evolved by involving more than 50 scientific partners spread from the north to the south of Italy in 30 cities coordinated by a volunteers' association, Frascati Scienza [2], constituted by citizens and researchers who strongly believe in the values of research and outreach.

The European Researchers' Night at European level had involved up to now more than 300 cities in all the member states with about 40 projects/year funded by a budget of 4 Meuro/year, that will become 6 Meuro/year from 2018.

Since 2006, the Italian project organised by Frascati Scienza not only has evolved in the number of partners and sites engaged, but also in typologies of contests, methodologies and objectives:

- 2006 – COME IN: lets share thousand emotions together
- 2007 – AGORÀ: enjoy beyond/with research/ers
- 2008 – EOS: Eyes On Researchers
- 2009 – SAY: Scientist Around Youth
- 2010 – BEST: focus on collaboration with European partners ERASMUS MC and CERN
- 2011 – BRAIN: Be in contact with Research And its Institutions Network for a night
- 2012 – RESPEcT: RESearchers - Pure Energy from Tip to Toe
- 2013 – TRAiL: Taste the ReseArchers" Life
- 2014/15 – DREAMS: Sustainability and Responsibility of Researchers and Science
- 2016/17 – MADE IN SCIENCE: Science as creating effective benefits for European citizens

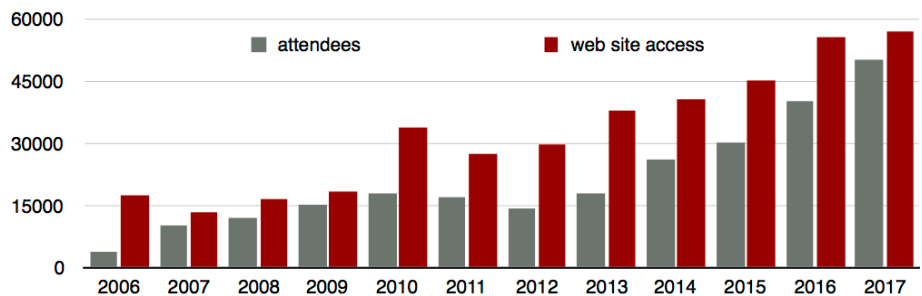

*Fig. 1 – European Researchers' Night attendees and www.frascatiscienza.it web site access during the awareness campaign period (about one month previous the manifestation)*

Fig. 1 shows the growth in terms of attendees and the accesses to the website (within the duration of the awareness campaign, usually one month around the NIGHT) in each year since 2006 to 2017. The last project, MADE IN SCINCE-2017, has consisted of more than 50.000 attendees, a media plan with 30 millions contacts, about 400 activities offered and 1800 researchers actively involved in the organization and realization of the scientific activities.

## 2 METHODOLOGY

### 2.1 Communication

Indeed, the awareness campaign has been one of the successful key points in the evolution of the project. Since 2009 the European Researchers' Night led by Frascati Scienza has started to employ professionals in communication activities to promote the relevant messages. A dedicated media plan has been designed and a specialized press office has been involved to promote the activities in the most important newspapers and TV channels at national level [3]. Between 2012-2014 it has been drafted the Frascati Scienza magazine, promoting outreach events throughout the whole year and large investments on social media channels (Facebook, Twitter, Instagram) has been developed. Moreover, Frascati Scienza has started to create several communication projects, like: radioscienza – aggregator of science news from research centres and researchers blogs [4], fisicast – podcast of physic contents [5], the science pills – funny videos with ten millions of visualization, awarded in shot movie festival and as sustainable non-profit advertising [6], #humansofERN, that emulating #humansofNY, to tell about citizens and scientists stories, and so on. At the same time, two independent projects, WIRE (Workshop Enterprises, Economy and Research) and Lazio Pulse [7], has been initiated to develop a dynamic ecosystem of public and private actors in the Lazio Region. These projects have increased the participation and collaboration of citizens in science activities.

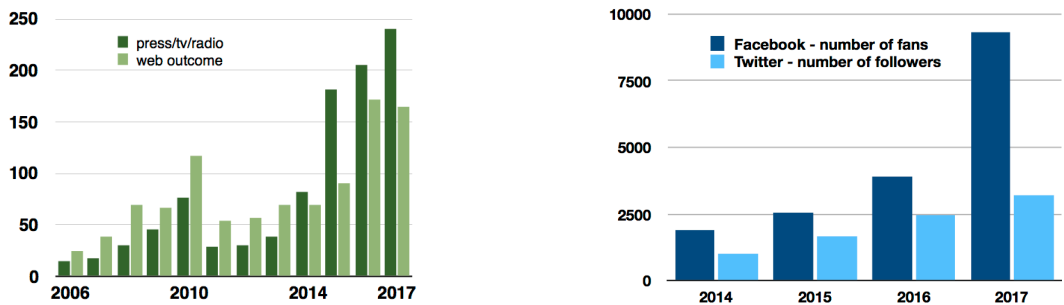

Fig 2 – left: press/tv/radio and web site outcome during the awareness campaign; right: growth of the Facebook' fans and Twitter' followers in the last years during the ERN

Figure 2 on the left shows the growth of traditional media since 2006, when first event has been realized. Figure 2 on the right shows the growth in terms of Facebook' fans and Twitter' followers Moreover, a dedicated newsletter with about 12000 contacts has been used to communicate science and innovation news, and to promote outreach events in the scientific area.

## 2.2 Activities

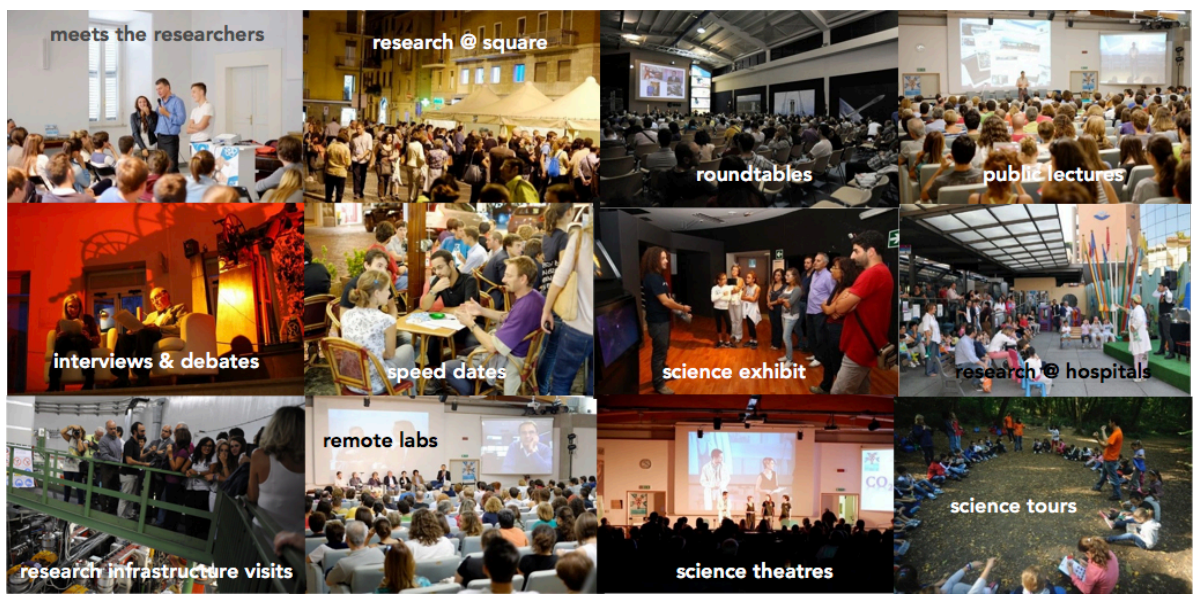

Fig. 3 – some picture of typical events

About 400 activities per year have been organized during the last ERNs; these are mainly divided in the following macro areas:

- *Hands-on activities:* small interactive experiments designed and presented by researchers, requiring the participation of the general public, with kids as the main target.
- *Workshops:* education activities focused on new trends in science and technology, such as open software, makers, etc., main target youngers.
- *Exhibitions:* showing the main research activities on going in the partners' scientific centres and cities, target general public.
- *Science shows:* theatre, movies, etc. mainly designed and performed by researchers showing science in a funny way, or aiming to present issues related to research, target general public.
- *Demonstrations:* live experiments, lectures, etc. addressed to demonstrate science principles, target general public.
- *Competitions:* previously and during the event competitions have been organized to attract the youngsters, main target young and kids.
- *Games and quizzes:* science games and quizzes aimed to involve people to participate in discovery and creativity activities, main target young and kids.

- *Informal talks with researchers:* researchers who in public spaces have improvised explanations and impromptu lessons on questions from the public.
- *Guided tours:* visits to the laboratories and research centers participating to the event, target general public.
- *Science trips:* visits of natural, historical, architectonical sites, etc., where the guide has been accompanied by a scientist who explained the applied scientific principles and technologies enabling discoveries, eg: the dating of the finds, the identification of false pictures, etc., etc.

In figure 3, a selection of pictures from the recent events are included, while in figure 4 the growth of events and partners during the recent ERN projects under the Frascati Scienza coordination are presented.

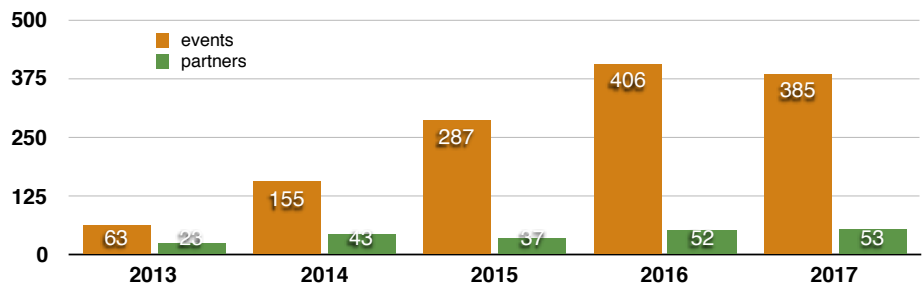

Fig. 4 – evolution of the activities and partners of ERN in recent years

Various outreach launch events have been carried out before the ERN, especially in spring, not only for public at large but also addressed to entrepreneurs and stakeholders: WIRE, Lazio Pulse, etc. [7]. A special event happened every year the Saturday preceding the NIGHT (flash-mob, hunting, science quits in the Rome center squares, etc.), and until the following Saturday many events are organized especially for the younger and schools to promote the European messages and motivate people to attend the clue events on Friday night.

The full programs of all the ERN' manifestation coordinated by Frascati Scienza since 2006 is available on the association website [8]

## 2.3 Impact

In order to evaluate the impact of the ERN, each year, a set of ex-ante and ex-post questionnaires have been filled in by the general public participating in the events, and a special questionnaire has been submitted to kids. Moreover, public games (such as the World Café, Trivia Night, etc.) have been organized to collect qualitative information about the impact of ERN. The main objectives of the survey were to:

- define the target audience reached during the European Researchers' Night;
- measure the increase of knowledge and promotion of the researcher's figure;
- draft the most possible concrete idea of the image of the researcher;
- understand the difference, between Italy and Europe, in terms of perception of the figure of the researcher and his role, especially in relation to employability and research funding;
- evaluate how much and which media have influenced the dissemination activities;
- understand the level of appreciation of the events.

In the following chapter, offering a comparison of data, the evolution of the results during the different years are presented. Moreover the recent perception of the ERN in the last event is shown, (29 September 2017) where a total of 2631 (617 ex-ante, 1.213 ex-post and 801 kids) questionnaires have been collected.

# 3 RESULTS

## 3.1 Audience and target

During the various projects some questions of the surveys have been improved to better understand the impact of the specific project or on requests of the European commission that year by year had to monitor different opinions. For this reason, we will report the data of the last survey, comparing mainly qualitatively with the results of the past.

In the last project the questionnaires were collected from all the cities participating to the project, although the majority have been gathered in Frascati (1.088), Rome (755), Milan (194) and Bari (79). The survey sampled about the 5% of the population attending the manifestation. Actually in Frascati this percentage grew at more than 10%, providing a reliable result.

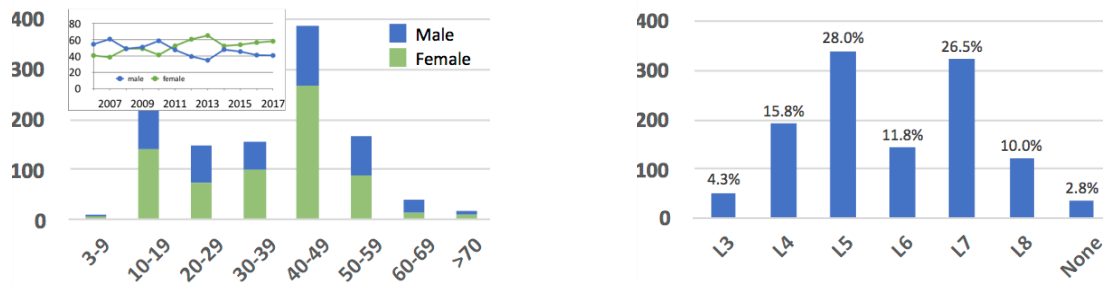

Fig. 5 – (ex-post questioners) left: gender (58% female, 41% male) and age of attendees (entries); left box age trend evolution (%); right: Qualification of attendees EQF Level (entries).

The highest percentage of respondents referred to the class of 30-49 years old (45%), probably the parents of kids' attendees. The majority was female, mostly in the recent years, as also supported by the interviews. This seems to be related to the greater interest of women to orient their children towards scientific interests and careers. Young people (3-29 years old) represent almost 37% of the total, which demonstrates the positive achievement of the target.

Qualification (Fig. 5) has been reported as EQF Level [9], in Italy equivalent to: (L3) primary school certification, (L4) middle school certification, (L5) high school certification, (L6) bachelor's degree, (L7) master degree, (L8) PhD or other equivalent qualification. The majority of attendees' had a High School degree (28%) and University degree (Bachelor's degree, 18%, and Master's Degree, 27%). The good level of qualification, higher then the Italian distribution [10], and the following qualitative information collected with the interviews, suggests that the attendees were not casual visitors, but people informed, motivated and interested in knowing more about science.

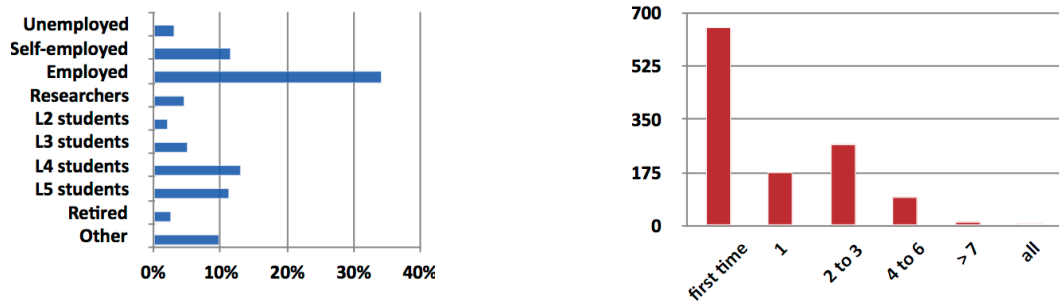

Fig. 6 – (ex-post questioners 2017) left: occupation (%); right: number of ERNs attended

The attendees (Fig. 6) were people with a job or studying, and the 46% declared to have participated to the previous editions, most of them have participated to more then one, enforcing the idea of a public strongly interested in science.

All this results and conclusions seem to be in good agreement with data collected in pervious edition [11]. Moreover, a detailed report year by year of the survey is available online [12].

## 3.2 Communication impact

The most effective communication channel (see fig. 7) to be aware of the ERN, as in previous editions, has been "Friends/Word of mouth" (37%), including "school/professors" among the other responses and "Social networks" (23%), communication channel that growing up year by year (fig 2 right). Moreover, Frascati Scienza website has resulted as a good media to know all the details about the European Researchers' Night (17%).

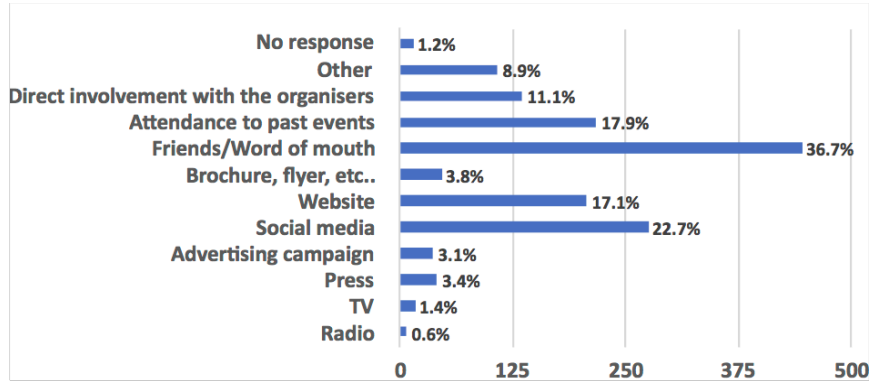

Fig. 7 – how attendees as been informed by the ERN manifestation (ex-post 2017)

Adding up all the contributions that fall under the awareness campaign (Radio, TV, Press, Advertising, social media, website, brochure) we can conclude that more then 52% of the people interviewed has been aware of the events from action designed for it, showing a very good impact of the media plan. This number it's not easy to compare with the past: not all the communication channel where monitored with the today accuracy or existed, like social networks. Anyhow we can assume that the same macro number stated at 28% in 2006 and has been growing up almost continuously over the years.

## 3.3 Perception

According to the data gathered, in Italy and in Europe, the role of research is considered very important for the development of Italy and Europe by almost all the interviewed (> 90%). But, only ~15% believes that research is adequately funded in Italy, while more than 75% of respondents is more confident regarding research funding at EU level. Both of the indicators have been stable in the recent years.

Most of the attendees (fig. 8 left) expected by the ERN a better understanding of the researchers' work and encouraging the interaction and dialogue with the researchers from the events. Among the "other" responses (not in figure), the most common comment has been: "to stimulate curiosity and interest in science among young people".

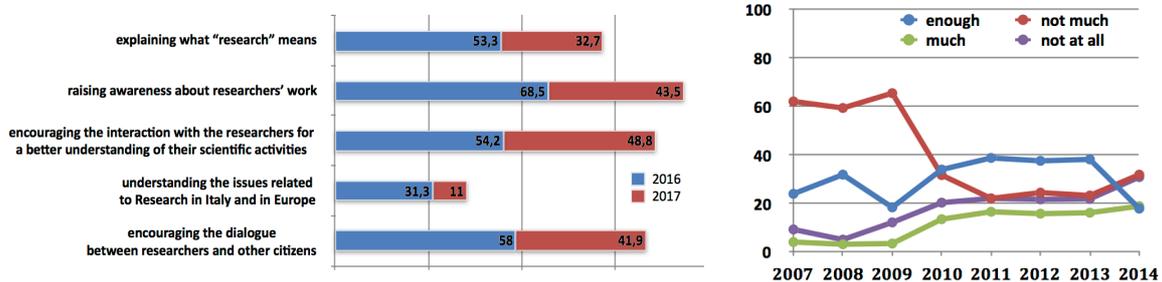

Figure 8 – left: opinion about the ERN objectives during MADE IN SCINCE project (%); right: how much the profession of researchers is felt part of the society (%)

In the past has been monitored also of how much the profession of researchers is felt part of the society (fig. 8: right) to motivate events like the ERN. In fact, almost all the respondents (> 90%) has supported that this kind of event is useful to promote the figure of researchers and to encourage young people to pursue a scientific career. The flexion in 2017 is due to change in the multiple choice of the survey to be more discriminating between the answers (fig. 9 on the left).

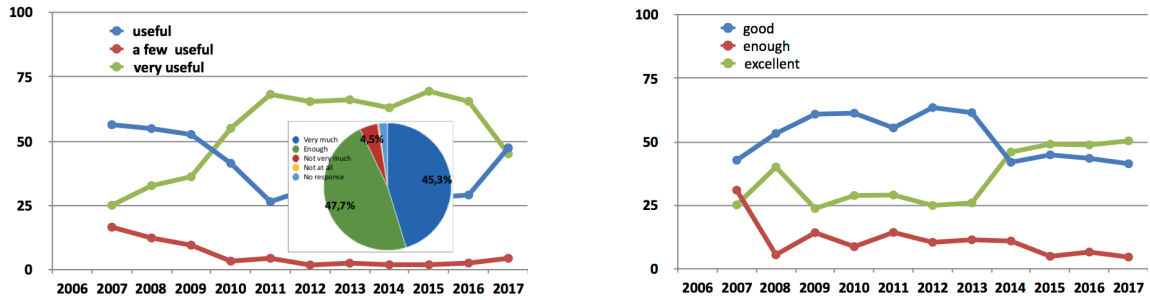

*Fig. 9 – left: trend showing the utility of ERN event to promote researchers' figure and a pie showing how much people think that ERN is useful for young scientific carrier (%);
right: efficiency of organization (%)*

The level of participants' satisfaction, regarding the efficiency of the organization, remains high and constant over the years (92%).

Finally, many participants left comments and suggestions for the future. Overall, the most of the participants were enthusiastic and thanked the organizers for the opportunity to talk with researchers. Very few respondents reported that researchers should talk in an easier way, calibrating their communication skills to the needs of the target. However, people appreciated the passion and the availability of researchers.

The interviewees suggested organizing more activities for kids in schools throughout the whole year, as only one Night is not enough.

## 3.4 Trivia Night

The qualitative perception of science and innovation, researchers' communication skills and impact of the ERN, seems to be aligned with the more general recent studies in Italy and Europe [13], [14], [15].

Besides the survey and the qualitative results collected, since 2014 games with the general public have been organized during the ERN to gather other qualitative information. In 2016 and 2017 manifestations a Trivia Night have been organized. It has consisted of an exciting quiz in team through which the staff of Frascati Scienza involved the general public to gather impressions and suggestions about the event and, at the same time, led it into the wonders of Science. The aim of the game was to bring out the public's views and analyze the strengths and weaknesses, opportunities and, above all, gathering ideas for the future editions of the ERN. In 2017 almost 100 people have attended, many adults between 25 and 60 years old, and children from 7 to 15 years old. The game relied on 14 questions divided in 7 sub-questions about scientific topics (Physics, Biology, Chemistry and Mathematics) and 7 questions about the ERN.

The answers about strengths has shown that participants considered the event as a unique opportunity to represent the deep sense of being European and to facilitate cooperation, sharing and communicating between countries, but also between different realities that many people think could not interact with different languages and competencies. Moreover, ERN brought science to people through a "simple language" without forgetting "irony" and "fun". Participants in the events recognized the value in "young and motivated researchers" and appreciated the decisive role in being a "stimulation to curiosity," "discovery," and "knowledge." The importance of both commitment and patience from the organizers has also been recognized.

Attendees have also been requested to score the following: the website accessibility, the awareness campaign, the wealth and originality of the program, the locations, the events. Events and program received a very high score, while a weakness has been underlined for the website and the awareness campaign of the manifestation. This does not seem in contrast to the success of the media plan, as the Trivia Night has been organized with citizens mainly from the city of Frascati. In fact, these people have highlighted the need for an awareness campaign more open and close to the territory.

Finally, other questions have been addressed to underline the expectation of attendees: "Would you love the night more if...? The request from many groups to increase activities, events and periods during the whole year with similar initiatives appears quite interesting. Another relevant point is the need to increase the interaction with laboratories. More and more ERN attendees seem ready to play an active role and underline the desire to have an ERN distributed on the territory, closer to schools, always richer in activities and experiments, promoter of meeting with great figures from the scientific world, longer (activities Saturday and Sunday), and more challenging (with revolutionary projects).

The most voted idea for a future slogan has been: *let's the thrilling of science upset you!*

## 4   CONCLUSIONS

The evolution of the European Researcher' Night project coordinated by Frascati Scienza, has shown continue growth of all the indicators during the recent years. The attendees are satisfied and motivated to participate in science, in particular young people, and, although they do not think to become scientists once grown up, they are enthusiastic to experience and play with science. The evolution of the filling respect to science communication and outreach activities appears to have evolved during the last twelve years of project evolution. During the first years, people were mainly driven by the curiosity to access research centers and meet the researchers; today they are much more educated and thirsty of knowledge. This indicates that the methodology adopted, although perfectible, seems to be successful as shown also from the more accurate report on the Italian and European perception of science and technology.

## ACKNOWLEDGEMENTS

We would like to thank Colette Renier, P.O. of the NIGHT since 2005, for the inexhaustible support during these years, the thousand of researchers involved and the institutions, associations and companies supporting and founding the manifestation: Regione Lazio, Comune di Frascati, ASI, CINECA, CREA, ESA-ESRIN, GARR, INAF, INFN, INGV, ISPRA, ISS, Sapienza Università di Roma, Sardegna Ricerche, Università di Cagliari, Università di Cassino, Università LUMSA di Roma e Palermo, Università di Parma, Università degli Studi di Roma "Tor Vergata", Università degli Studi Roma Tre, Università di Sassari, Astronomitaly, Associazione Tuscolana di Astronomia, Explora, G.Eco, Ludis, Osservatorio astronomico di Gorga, Sotacarbo, Associazione Eta Carinae, Cicap Lazio, Consorzio di Ricerca Hypatia, Engineering, Fondazione GAL Hassin di Isnello, GEA, Giornalisti nell'Erba, FVA New Media Research, ICBSA, Istituto Nazionale Tumori Regina Elena e Istituto Dermatologico San Gallicano – IRCCS Roma, Matita Entertainment, IRCCS Lazzaro Spallanzani, Museo Geopaleontologico "Ardito Desio" di Rocca di Cave, Osservatorio Malattie Rare, Museo Tuscolano delle Scuderie Aldobrandini, STS Multiservizi, Ass. Speak Science, Ass. ScienzImpresa, Tecnoscienza.it srl, The Document Foundation, Unitelma Sapienza, Università della Tuscia, Istat Sicilia, Hamamatsu Photonics Italia S.r.l., Accatagliato, Associazione culturale Arte e Scienza, Associazione culturale Chi sarà di Scena, Associazione Amici di Frascati, Res Castelli Romani, MaCSIS, AGET Italia, Istituto salesiano di Villa Sora, Pro Loco Frascati 2009.

The European Researchers' Night project is funded by the European Commission under the Marie Skłodowska-Curie actions (Grant Agreement No. 722952)